\documentstyle[aps,prb,preprint]{revtex}
\begin{document}
\draft
\title{Microwave Photoresistance Measurements of Magneto-excitations near a 2D
Fermi Surface}
\author{M. A. Zudov, R. R. Du}
\address{Department of Physics, University of Utah, Salt Lake City, UT 84112}
\author{J. A. Simmons, J. L. Reno}
\address{Sandia National Laboratories, Albuquerque, NM 87185}
\date{\today }
\maketitle

\begin{abstract}
We report the detection of magneto-excitations, {\em i.e.} the cyclotron
(CR) and magnetoplasmon (MPR) resonances, near the Fermi surface of a
high-mobility two-dimensional electron system (2DES), by microwave
photoresistance measurements. We observe large amplitude photoresistance
oscillations originating from higher order CR, {\em i.e.} transitions
between non-adjacent Landau levels. Such transitions are drastically
enhanced in low magnetic field as compared to those previously known in the
high field limit. The scattering time of the CR is found to be nearly one
order of magnitude larger than that of the Shubnikov-de Haas oscillations.
Finally, distinct photoresistance peaks are observed in addition to the CR
features. They are identified as resonances of the low-frequency MP modes at
a cut-off wavelength, determined by the width of the 2DES sample.
\end{abstract}


Low-energy collective excitations in a quantum Hall system both in the bulk
and at the edges of the quantum fluid are of intense current interest\cite
{ref1}. In particular their electrodynamical response to a high frequency $%
(\omega )$, finite wavevector $(q)$ probing field is believed to yield
information complementary to dc transport measurements, for example on the
internal structure of the edge states\cite{ref2}.

The relevant energy of these excitations is dictated by the Coulomb
interaction, and is typically in the range\cite{ref1} of $1$ to $10K$.
However, established experimental techniques have largely been limited to
the high-energy regime where the cyclotron energy $\hbar \omega _c$ is
comparable to the Fermi energy $\varepsilon _F$ of the two-dimensional
electron system (2DES). For example, dynamical transport measurements of the
cyclotron (CR)\cite{ref3,ref4} and magnetoplasmon resonances (MPR)\cite
{ref5,ref6,ref7} of a 2DES have been obtained by employing far-infrared
(FIR) spectroscopy\cite{ref8} down to about $10$ $cm^{-1}(15K)$. Only
recently have spectroscopic measurements in the low-energy regime, {\em i.e.}%
, millimeterwave frequencies, become available\cite{ref9,ref10}.

In this work we report the detection of CR and MPR near the Fermi surface $%
(\hbar \omega _c\ll \varepsilon _F\approx 80K)$ of a high-mobility 2DES by
microwave photoresistance\cite{ref11} measurements. Unexpectedly, in a
high-mobility GaAs-AlGaAs 2DES subject to a {\em weak} magnetic field, we
observe giant photoresistance oscillations associated with higher order CR, 
{\em i.e.}, transitions between non-adjacent Landau levels (LL). Such
transitions have previously been seen in CR experiments in the high magnetic
field $(B)$ regime\cite{ref3} and have been explained via an analysis of
short-range scattering potentials\cite{ref12}. In the high $B$ limit,
however, the amplitude of even the second order CR is rather small. The
dramatic enhancement of the higher order CR observed here is attributed to
the mixing of many LLs in a weak $B$ field\cite{ref13}. We have also
observed distinct photoresistance peaks associated with the dispersion of
the MP modes. These peaks occur at a distinct cut-off wavelength of the MP
mode, given naturally by the width of the sample.

Our samples are standard Hall bars lithographically defined (width $200\mu m$%
) and etched from modulation-doped MBE GaAs-AlGaAs heterostructures. While
similar results have been obtained from a variety of samples, data from only
one sample are presented here unless otherwise indicated. The sample has an
electron density $n$ $=2.0\times 10^{11}cm^{-2}$, and a mobility $\mu
\approx 3.0\times 10^6cm^2/Vs$, obtained by a brief illumination from a red
light-emitting diode. The distance between the electrons and the Si doping
layer is $d_s\approx 700\AA $. Coherent, linearly polarized, millimeter wave
radiation from a solid state source of tunable frequency $f=\omega /2\pi $
from $30$ to $150GHz$ $(1.5$ to $7.5K)$ is guided via an over-sized
waveguide to the sample. The sample is immersed in the $^3He$ coolant of a
sorption-pumped $^3He$ refrigerator at a constant temperature $(T)$ ranging
from $0.4$ to $1.8K$. The output power of the source is from $10$ to $50mW$;
between the source and the waveguide an attenuator is placed for varying the
incident power to the sample. $B$ is applied perpendicular to the sample
surface. We measure the magnetoresistance employing a standard low frequency 
$(3$ to $7Hz)$ lock-in technique.

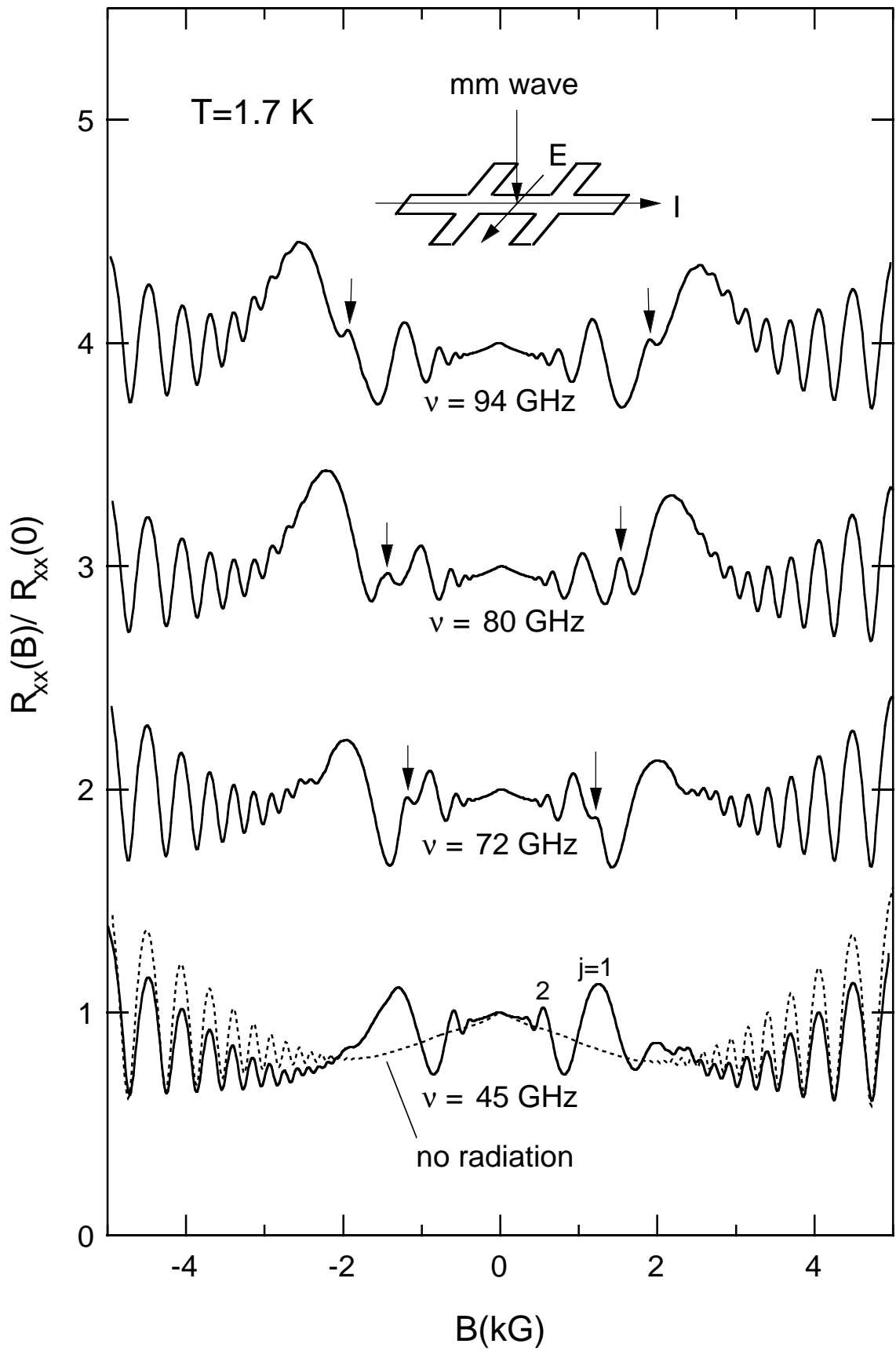
\begin{figure}[tbp]
\caption{Magnetoresistance in a weak magnetic field under millimeter wave
illumination at several fixed frequencies $f =\omega /2\pi $, measured at a
constant temperature of $T=1.7K$. The data exhibits large amplitude
photoresistance oscillations for all frequencies. A distinct, additional
peak (arrow) emerges at high frequencies. The inset depicts the microwave $E$
field polarization with respect to the sample.}
\label{fig1}
\end{figure}

In FIG.\ref{fig1} we show the normalized magnetoresistance $R_{xx}^\omega
(B)/R_{xx}^\omega (0)$ under microwave illumination for several values of $%
\omega $. The $R_{xx}^\omega $ is measured by sweeping $B$ while $\omega $
and $T$ are held constant. Without illumination (dotted line), the sample
starts to exhibit normal Shubnikov-de Haas (SdH) oscillations at an on-set
field $B_{SdH}\sim 2.0kG$. New, large amplitude resistance oscillations
appear under illumination, arising from a correction to the
magnetoresistance owing to the resonant absorption of microwave radiation%
\cite{ref13}. The correction, or the photoresistance $\Delta R_{xx}^\omega
(B)=R_{xx}^\omega (B)-R_{xx}^0(B)$ generally alternates in sign with
increasing $B$ so as to form an oscillatory structure which is roughly
periodic in $1/B$ (see {\em e.g.}, the trace at $45GHz$). With increasing $%
\omega $, such structure shifts to higher $B$ in an orderly fashion, but
largely remains limited to a low $B$ range where the SdH is absent.

The $B$ positions of the $\Delta R_{xx}^\omega (B)$ peaks are seen to
satisfy a resonance condition in which the microwave quantum equals an {\em %
integer\ multiple} $j$ of the cyclotron energy $\hbar \omega _c=\hbar
eB/m^{*}$, where $m^{*}$ is the electron effective mass, 
\begin{equation}
hf=j\cdot \hbar \omega _c\text{, or }j=\omega /\omega _c\text{.}  \label{eq3}
\end{equation}
Thus, for $j>1$, the structure is due to microwave-induced transitions
between non-adjacent LLs, or higher-order CR. More surprisingly yet, a
distinct $\Delta R_{xx}^\omega $ peak emerges for frequencies between
roughly $60GHz$ and $100GHz$, with an amplitude and a width in $B$
comparable to those of the CR features. The $B$ field position of this
additional peak shifts from the $j=2$ towards the $j=1$ CR peak as $\omega $
increases, with the feature eventually merging with the $j=1$ peak. This is
the MP resonance to which we will return.

While higher-order CR has previously been observed, the low $B$ field regime
of this work produces a strong mixing of multiple LLs. This results in a
dramatic enhancement of the amplitudes for higher-order transitions, as
discussed in Ref.13. In fact, under our low $B$ condition the
photoresistance is caused by the lowest-order corrections to the
conductivity in a microwave field, which is given by\cite{ref13}: 
\begin{equation}
\Delta \sigma _{xx}(\omega )\propto \cos (2\pi \omega /\omega _c)\cdot \exp
(-2\pi /\omega _c\tau )\text{.}  \label{eq4}
\end{equation}

Several comments are appropriate here. While such photoresistance
oscillations may resemble those of SdH, they differ in the following crucial
aspects. First, the period is controlled by the ratio $\omega /\omega _c$
rather than $\varepsilon _F/\hbar \omega _c$, indicating a purely classical
effect. The absence of the Fermi energy $\varepsilon _F$ in Eqn.\ref{eq4}
also implies that the effect does not depend on the electron density. This
is supported by similar data obtained from other samples of differing
density. In addition the oscillation is expected to have very weak $T$
dependence within a certain temperature range. Indeed the amplitude of the
oscillation does not change significantly from $0.4$ to $1.8K$, whereas the
SdH exhibits very strong $T$ dependence. Finally, the damping exponential
contains a factor of $2$ as compared to the usual Dingle factor in the SdH
formalism\cite{ref14}.

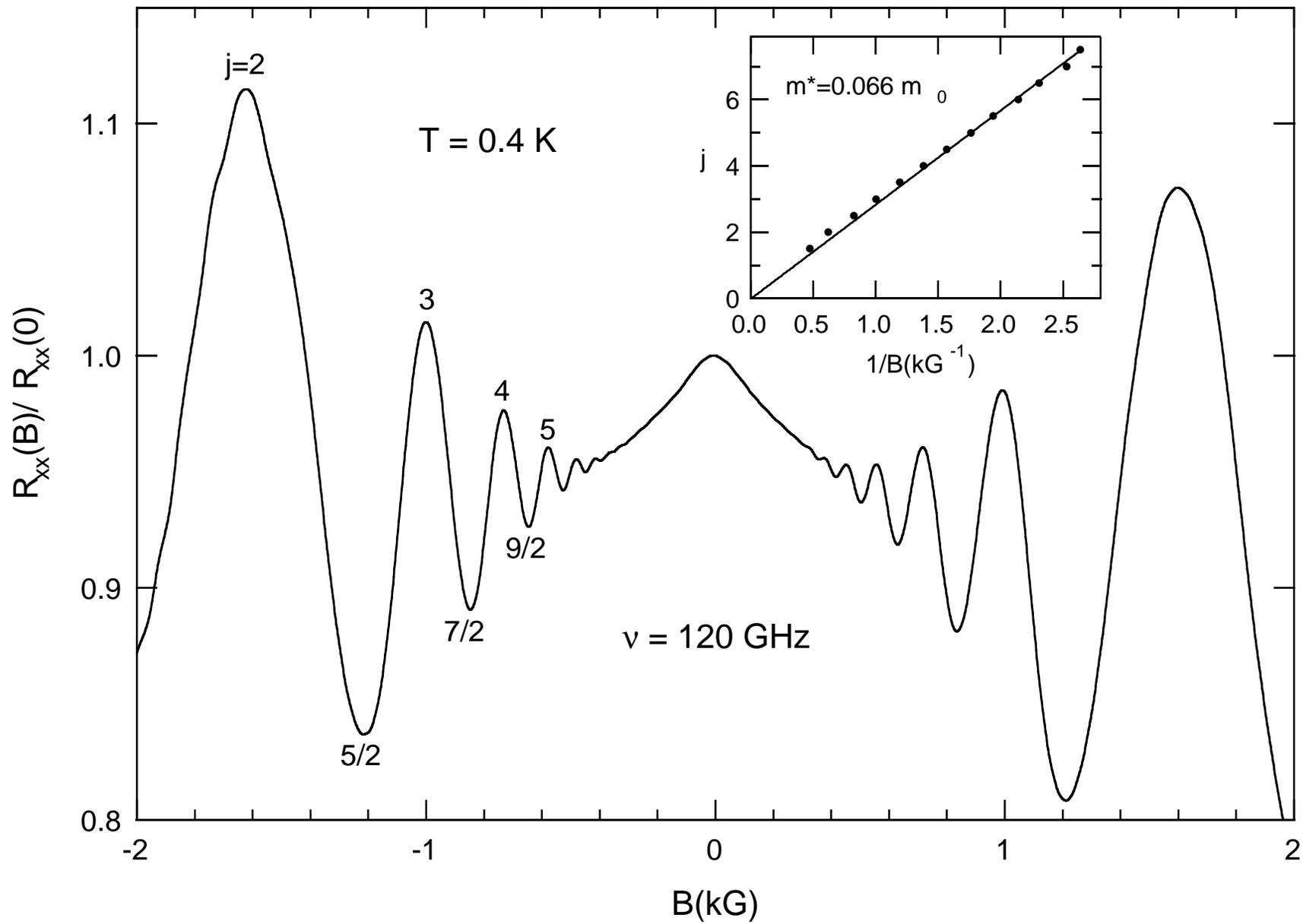
\begin{figure}[tbp]
\caption{Photoresistance at $f =120GHz$ showing multiple extrema persisting
to $j=7$. The inset is a plot of the order index $j$ against the inverse
magnetic field position of resistance extrema, yielding an electron mass of $%
m^{*}=0.066m_0$.}
\label{fig2}
\end{figure}

Having identified the higher-order CR from the photoresistance, we have
further determined respectively the mass $m^{*}$and the scattering time $%
\tau _{CR}$. From Eqn.\ref{eq3}, $j=\omega /\omega _c=\omega m^{*}/eB$,
therefore for a fixed $\omega $ we may plot the multiplicity index $j$
against the inverse magnetic field $1/B$, to obtain the value of $m^{*}$.
Such a plot is shown in FIG.\ref{fig2} for the data at $120GHz$ where the
oscillations are seen to persist to as high as $j=7$. We include here both
the integer $j=2,3,...$ (maxima in photoresistance ) and the half integer $%
j=3/2,5/2,7/2,...$ (minima), since both are described by Eqn.\ref{eq4}. The
effective electron mass yielded is $m^{*}\approx 0.066$ (in units of the
free electron mass), close to the known value $0.068$ of the band electron
mass in GaAs.

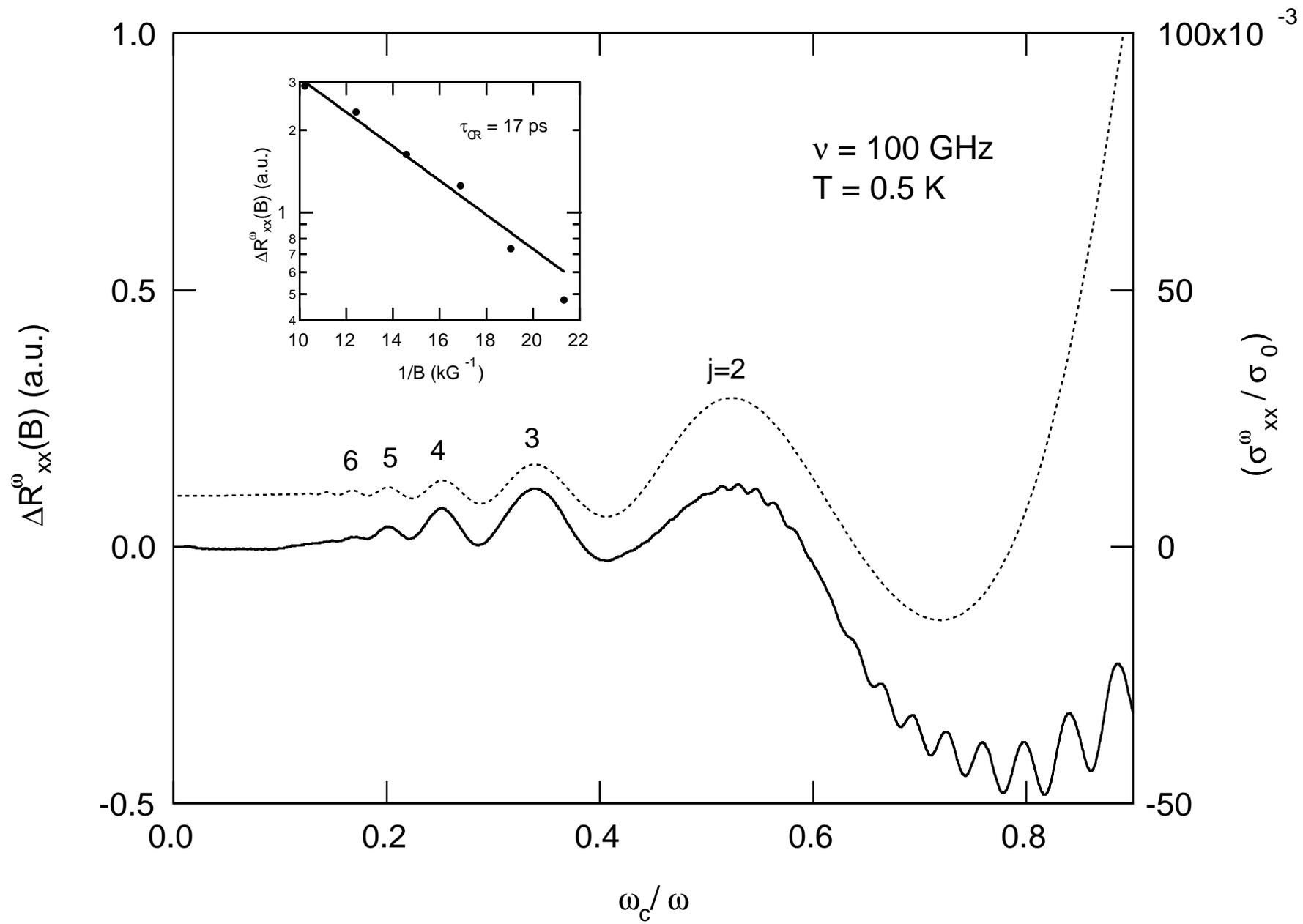
\begin{figure}[tbp]
\caption{Comparison of photoresistance oscillations from experiment (solid
line) with the dissipative conductivity from the theory of Ref.13(dotted
line). The theoretical curve assumes $\omega \tau _{CR} =10$, corresponding
to a scattering time of $16$ $ps$. Inset: plot of the oscillation amplitude
vs. the inverse of the magnetic field, consistent with a scattering time $%
\tau _{CR}\approx 16$ $ps$ obtained from Eqn.\ref{eq4}.}
\label{fig3}
\end{figure}

In order to estimate $\tau _{CR}$, we compare the model of Ref.13 to the
data, as shown in FIG.\ref{fig3}. This model takes into account the
scattering corrections for each harmonic, and therefore is improved in
details over Eqn.\ref{eq4}. The dotted line represents the calculated
conductivity $\sigma _{xx}^\omega /\sigma _{0\text{ }}^\omega $(and hence
the absorption) with a scattering parameter $\omega \tau _{CR}$. As shown
here, the data of the photoresistance (solid line) and the $\omega \tau
_{CR}=10$ curve agree with each other not only in periodicity but in the
relative amplitude of the harmonics as well, yielding\cite{ref14} a
scattering time $\tau _{CR}\approx 16ps$. This value compares well with that
obtained from the simpler Eqn.\ref{eq4} (shown in the inset to FIG.\ref{fig3}%
).

The SdH scattering time $\tau _{SdH}\sim 1ps$ in the same sample determined
by a Dingle plot (not shown) is, however, much smaller than $\tau _{CR}$. We
also estimate the transport scattering time from the dc mobility $\mu =e\tau
_t/m^{*}$ to be at $\tau _t\approx 100ps$. Comparing scattering times in
these three regimes we speculate that the CR oscillations observed here may
not be sensitive to certain properties such as the density inhomogeneity in
a 2DES. It is well known\cite{ref15} that the density inhomogeneity
contributes to an underestimate of the experimental value of $\tau _{SdH}$.
On the other hand, there is no existing theory which deals with
photoconductivity in realistic samples. We note also that a very large CR
scattering time, even larger than that of the transport scattering time, has
been observed in the extreme magnetic quantum limit in a GaAs-AlGaAs system%
\cite{ref4}.

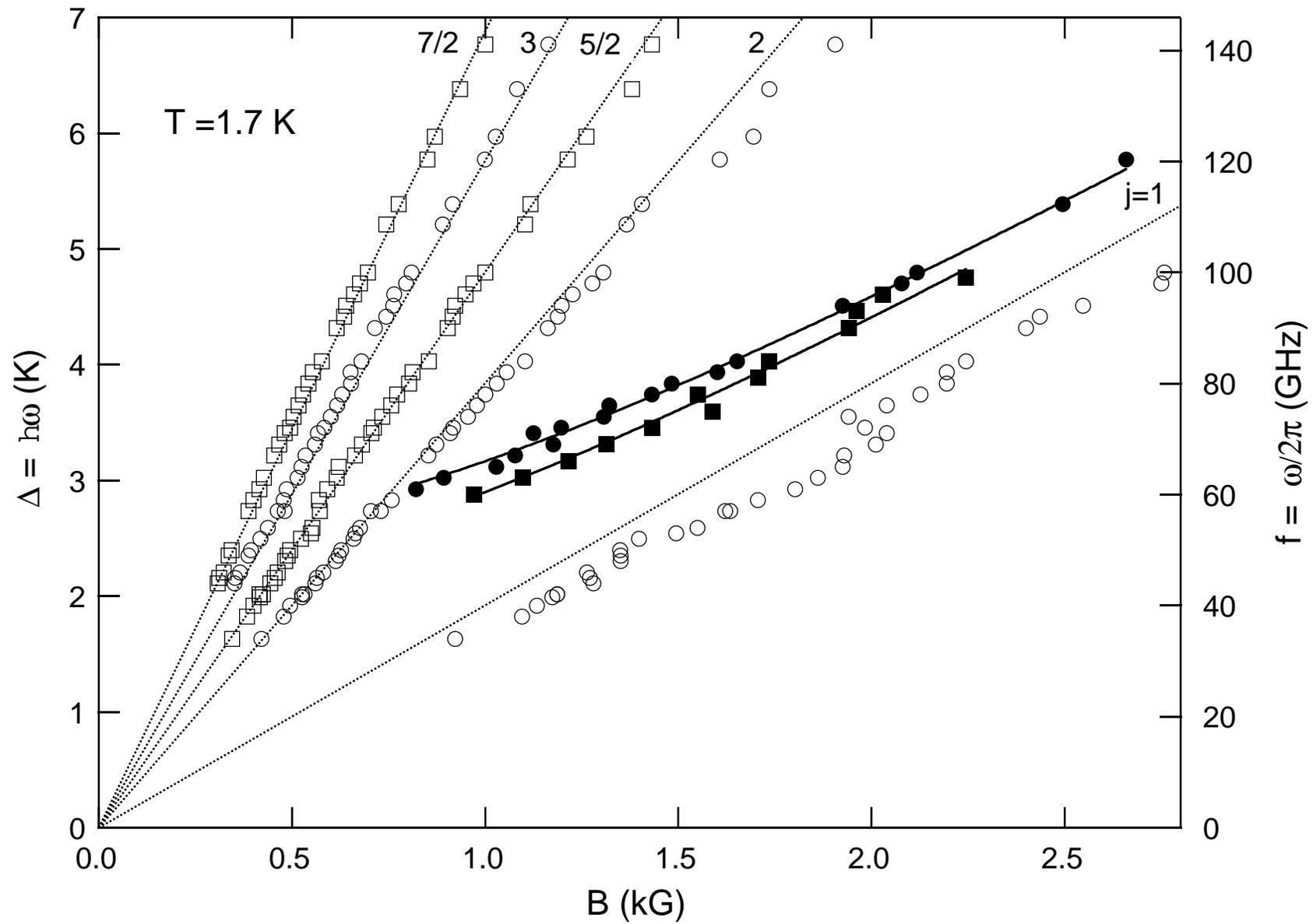
\begin{figure}[tbp]
\caption{Resonant energy $\Delta =\hbar \omega $ (in Kelvin) versus magnetic
field $B$ for resistance extrema $j=1,2,3$ (open circle) and $j=5/2,7/2$
(open square). The dotted lines represent the value of $0.068m_0$ known for
the band electron mass in GaAs. The filled circle (filled square) is the
data of the magnetoplasmon resonance in the sample having an electron
density $n=2.0\times 10^{11}cm^{-2}$ ($1.7\times 10^{11}cm^{-2}$) and a Hall
bar width $W=195\mu m$ ($200\mu m$). The solid lines are the fits of the
magnetoplasmon dispersion to each sample, as discussed in the text. }
\label{fig4}
\end{figure}

We have further confirmed our CR results over a wide range of $\omega $. In
accordance with Eqn.\ref{eq3}, the energy required for higher-order CR, for
a set of $integer$ $j$, is expected to form a fan diagram against the $B$
field position of each resistance maximum. In FIG.\ref{fig4}, we plot our
data in just such a diagram covering our $\omega $ range. The open symbols
represent the data for each branch $j=1,2,3$, and the dotted lines represent
the fans calculated using a value of $0.068$ as the electron mass. The
higher-order extrema ($j=2,5/2,3,...$) data are described well by Eqn.\ref
{eq3}, over a wide range of frequencies. However, the fundamental CR, i.e.,
the $j=1$ transition energy, is roughly lower by $10\%$ as compared to the
calculated cyclotron gap. The origin of the discrepancy between the
calculated value and the data at $j=1$ is unclear at this point, especially
in light of the good agreement obtained for the higher-order branches.

We devote the rest of the paper to a discussion of the magnetoplasmon. The
MPR data, both for the same sample ($n$ $=2.0\times 10^{11}cm^{-2}$, filled
circles) and for an additional sample having a lower density ($n$ $%
=1.7\times 10^{11}cm^{-2}$, filled squares), are shown in FIG.\ref{fig4}.
The existence of these peaks in our low magnetic field measurements is
rather puzzling at first glance. On the one hand, the magnetic field
dependence of the resonance energy resembles, when plotted in this fashion,
the finite-wavevector MP that has been seen\cite{ref7} in a
grating-modulated 2DES at high $B$ fields. On the other hand it is necessary
to introduce a finite $q$ in the 2DES for the MP resonance to take place.

In the simplest approximation, which ignores non-local interactions, the
dispersion of a long wavelength ($\omega /c<q\ll q_F$, where $q_F$ is the
Fermi wavevector) plasmon is known to be\cite{ref5,ref6,ref7}: 
\begin{equation}
\omega _P^2(q)=\frac{ne^2q}{2\epsilon \epsilon _0m^{*}}  \label{eq1}
\end{equation}

where $\epsilon $ is the effective dielectric function of the surrounding
media. In the presence of a perpendicular magnetic field, the plasmon has a
low cut-off frequency given by the cyclotron frequency $\omega _c=eB/m^{*}$
and the coupled cyclotron-plasmon modes or the MP dispersion then becomes 
\begin{equation}
\omega _{MP}=\sqrt{\omega _c^2+\omega _P^2}\text{ }.  \label{eq2}
\end{equation}
The MP dispersion dictates that the MP resonances can be observed only by
coupling to a finite momentum transfer $q$, usually via a spatially
modulated radiation field or electron density\cite{ref5,ref6,ref7}. The
observation of MP absorption at a specific $\omega $ in an un-patterned 2DES
implies, then, that a finite $q$ must be selected in the process. It turns
out that what we have observed are just the resonances of the low-frequency
MP modes at a cut-off wavelength given by the width of the 2DES sample.

Empirically we fit the MP dispersion (Eqns.\ref{eq1},\ref{eq2}) to our data
to find the value of $q$, using a mass $m^{*}\approx 0.068$ and a dielectric
constant $\epsilon =12.8$. Simple fits result in, respectively, $2\pi
/q\approx 204\mu m$ $($for the $n=2.0\times 10^{11}cm^{-2}$ sample$)$ and $%
229\mu m$ $(1.7\times 10^{11}cm^{-2})$. These values compare quite well with
the lithographic width $(\sim 200\mu m)$ of both Hall bars, considering that
several factors contribute corrections to the idealized 2D plasmon
dispersion given by Eqn.\ref{eq1}. In the long-wavelength limit, the finite
thickness of the 2D electron layer and the uncertainties of the dielectric
function $\epsilon $ are thought to be the main corrections\cite{ref7}. The
former in particular will soften the plasmon frequency, therefore rendering
an overestimation of the width. Depletion of the electrons near the edge
will, of course, reduce the effective width, but such an effect is expected
to be negligibly small here.

It is then convincing that a low-frequency branch of MP is excited by
microwave fields in our samples, and detected by our photoresistance
measurements. We further note that since polarization of the microwave $E$
field is perpendicular to the length of the Hall bar (inset FIG.\ref{fig1}),
the MP observed are transverse to the current. The MP modes observed here
can then be viewed as the standing waves in a 2D waveguide defined by the
Hall bar, and these are just the low-frequency cut-off modes. In fact, we
have also observed higher-order MP modes associated with integer fractions
of the sample width.

To summarize, we have identified the oscillatory photoresistance as the CR
of electrons in a weak magnetic field. We emphasize that the exceptionally
strong higher-order harmonic absorption observed here is due to the presence
of short-range scatterers, ubiquitous in even very clean GaAs-AlGaAs
heterostructures, and that the presence of higher LLs favors large
amplitudes for the higher-harmonic absorptions. The scattering time in this
regime appears, however, an order of magnitude larger than that determined
by the Dingle plot in the SdH regime. This observation may be attributed to
the different role played by inhomogeneity in these two regimes. We have
also identified distinct resonances of the low-frequency magnetoplasmon
modes at a cut-off wavelength, determined by the width of the sample. This
work thus establishes a method for the detection of low-energy
magneto-excitations close to the Fermi surface of a 2DES. It is our hope
that this experimental technique of microwave resonant photoconductive
spectroscopy can be readily extended to a number of low-dimensional electron
systems.

We are grateful to D. C. Tsui for encouragement and valuable discussions,
and for comments on the manuscript. We thank S. J. Allen Jr. for helpful
conversations on experiments and especially M. E. Raikh for extensive
communications throughout this work. The work is supported in part by NSF
grant DMR-9705521 (M.A.Z. and R.R.D.). R. R. D. also acknowledges a Alfrad
P. Sloan Research Fellowship. The work at Sandia is supported by the U.S.
DOE under contract DE-AC04-94AL85000.

\begin{references}
\bibitem{ref14}  A quantitative fit to the absolute amplitude of the
photoresistance oscillations is not attempted since, as has been pointed out
elsewhere\cite{ref11}, details of the photoconductive processes remain
inadequately understood.
\end{references}
\end{document}